\documentclass[11pt,a4paper]{article}
\usepackage{latexsym}
\usepackage{amsmath}
\usepackage{amsthm}
\newtheorem{theorem}{Theorem}[section]

\usepackage{graphicx}
\usepackage{epstopdf}
\usepackage[round]{natbib}
\usepackage{color}
\usepackage{pdfpages}
\usepackage[margin=1in]{geometry}
\usepackage{authblk}
\newcommand{\be}{\begin{equation}}
\newcommand{\ee}{\end{equation}}
\newcommand{\bea}{\begin{eqnarray}}
\newcommand{\eea}{\end{eqnarray}}
\newcommand{\mb}{\mathbf}

\newcommand{\bsb}{\boldsymbol}

\begin{document}
\newcommand{\vect}[1]{\boldsymbol{#1}}
\title{Mixture cure semiparametric additive hazard models under partly interval censoring 
-- a penalized likelihood approach
}
\author[1]{Jinqing Li \footnote{To whom correspondence should be addressed: actuaryljq@uibe.edu.cn}}
\author[2]{Jun Ma }

\affil[1]{Department of Statistics and Actuarial Studies,
School of Insurance and Economics,
University of International Business and Economics, China}
\affil[2]{Depart of Mathematics and Statistics, Macquarie University, Sydney, Australia}
\date{}
\maketitle
\begin{abstract}
Survival analysis can sometimes involve individuals who will not experience the event of interest, forming what is known as the ``cured group''. Identifying such individuals is not always possible beforehand, as they provide only right-censored data. Ignoring the presence of the cured group can introduce bias in the final model. This paper presents a method for estimating a semiparametric additive hazards model that accounts for the cured fraction. Unlike regression coefficients in a hazard ratio model, those in an additive hazard model measure hazard differences. The proposed method uses a primal-dual interior point algorithm to obtain constrained maximum penalized likelihood estimates of the model parameters, including the regression coefficients and the baseline hazard, subject to certain non-negativity constraints.
\\[2ex]
\end{abstract}

\section{Introduction} \label{sec:intro}
When analyzing survival times, it is common to encounter situations where the event of interest does not occur for some individuals. 
For example, in the context of cancer recurrence after treatment of thin melanoma (defined as tumor Breslow thickness less than 1mm), since the prognosis for thin melanoma patients is generally favorable, there is a high probability that a patient will not experience cancer recurrence after treatment \citep{WeJuSe22}. The subgroup of individuals who do not experience the event-of-interest is known as the cured fraction. Individuals in the cured fraction are generally not known a priori, and they will only contribute right-censored times.
When observed event times contain both susceptible and cured fractions, it is difficult to classify a right-censored time into one of these two groups.
Analyzing such a data set without taking the cured fraction into consideration (i.e. treating all the right-censored times as true censoring times of the event) may lead to biased parameter estimates (e.g, \cite{Farewell82, sy_estimation_2000, WeJuSe22}).

A common approach, when a cured fraction presents, is to adopt a
mixture cure Cox model \citep{Farewell82}, 
where a Cox model is assumed for the susceptible fraction and a logistic model 
for the cured fraction.
However, in certain circumstances, people may be interested in an alternative to a mixture cure Cox model. This paper studies a particular alternative, namely the mixture cure additive hazards (AH) model.

The AH model, which was first introduced by \cite{Aalen89} and further developed by \citet{LinYin94}, is a semiparametric model represented by equation (\ref{AH model}) below. This model assumes that the hazard function of each subject is the sum of a non-parametric baseline hazard and a parametric linear predictor. In an AH model, the regression coefficients represent hazard differences, whereas in a Cox model, the coefficients are related to 
hazard ratios.

When combined with a cure fraction using a mixture distribution, the AH model becomes 
a mixture cure AH model, as described by \cite{Wang12}.

There are various methods available in the literature for fitting mixture cure AH models, such as those discussed in \cite{YaHeLuSo22} and \cite{Wang12}. However, these methods suffer from certain issues. 
Firstly, they do not take into account the non-negativity constraints that apply to both the baseline and individual hazard functions. Although the non-negativity is considered for the baseline hazard, these methods overlook the fact that the hazard of each individual must also be non-negative.
Secondly, their estimation procedures rely heavily on the expectation-maximization (EM) method. 
While EM is a stable algorithm, it does not provide directly the covariance matrix for the estimates, as noted by \cite{Louis82}. Several researchers have suggested computationally intensive methods, such as bootstrapping, to compute the standard errors. However, a bootstrap method may not be efficient for survival regression models when the sample size is large.

We would like to point out that some of the non-negativity constraints 
can be active, especially when the knots used for approximating the baseline hazard are not selected carefully. 
Ignoring active constraints can cause unpleasant consequences, such as negative asymptotic variances for some of the estimated parameters as discussed in \cite{MaCoHeIa21}. 
A technique has been developed in \cite{MaCoHeIa21} to handle active constraints. A similar technique will be adopted in this paper 
to develop the asymptotic variance result for the model under consideration.

This paper focuses on fitting semiparametric mixture cure AH models where the event data from the susceptible fraction are allowed to be partly interval-censored, where the definition of partly interval censoring can be found in \cite{Kim03}. Briefly, partly interval-censored survival data can include event times as well as left, right and interval censoring times. We estimate the model parameters by maximizing a penalized log-likelihood function, with the penalty function used for two main purposes: (i) to smooth the non-parametric baseline hazard, and (ii) to reduce the requirement for the optimal number and location of knots. The optimization process must also consider the non-negativity constraints that apply to the baseline hazard and the hazards for all individuals.

To find the constrained maximum penalized likelihood (MPL) estimates of both the regression coefficients and the baseline hazard, we adopt a primal-dual interior point algorithm as described in \cite{Wright97}. Additionally, we provide an asymptotic covariance matrix that is adjusted for active constraints. This allows us to calculate standard errors of the estimates without relying on computationally intensive methods.

The remainder of this paper is structured as follows. Section \ref{sec:mple} presents the formulation of the constrained MPL estimation problem. In Section \ref{sec:algoInteriorPoint}, we provide a summary of the primal-dual interior-point algorithm and detailed information on this algorithm is provided in the Supplementary Materials. We then explain in Section \ref{sec:smpa} an optimal smoothing parameter selection method using a marginal likelihood.
In Section \ref{sec:asym}, we establish the consistency and asymptotic normality of the proposed MPL estimates and in Section \ref{sec:resu}, we present the results of a simulation study and an application of our method to a thin melanoma dataset. Finally, we provide some concluding remarks in Section \ref{sec:conc}.

\section{Model and likelihood function} \label{sec:mple} 
For individual $i$, there are two possibilities: either this person is a member of the susceptible (non-cured) fraction (so that the event-of-interest will eventually occur) or a member of the cured fraction (so that the event will never occur). To accommodate these two sub-populations we define an indicator $U_i$ such that $U_i=1$ if individual $i$ is in the susceptible group and $U_i=0$ for otherwise. We use $T_{i}$ to denote a pseudo event time for individual $i$. Thus, $T_i\, |\, U_i = 1$ represents the event time of interest since $i$ is in the susceptible fraction, $T_i\, |\, U_i=0$ represents the end of follow-up time for a cured individual. Note that for the latter case $T_i$ will be recorded as a right censoring time. For $i$, its $U_i$ value is unknown only when $T_i$ is right-censored; for all other cases we have 
$U_i=1$.

In this paper we consider general partly interval censoring for the observed survival times. This means the observed times can include event times or left, right and interval censoring times. In this context, the observed survival times can be conveniently denoted by $(t_i^L, t_i^R)$ for $i = 1, \ldots, n$. Clearly, when $t_i^L = t_i^R=t_i$ we have an event time, $(0, t_i^R)$ indicates $t_i^R$ is a left censoring time, $(t_i^L, \infty)$ means $t_i^L$ is a right censoring time and all other cases correspond to interval censoring times.

We use $\pi(\mb{z}_i)$ to denote the probability of $U_i=1$, where $\mb{z}_i = (z_{i1}, \ldots, z_{iq})^\top$ is a vector of covariates 
to model $\pi(\mb{z}_i)$. We adopt a logistic model for $\pi(\mb{z}_i)$ throughout this paper. On the other hand, we consider a semiparametric AH model (see \cite{LinYin94} and \cite{LiMa19})
for event times of the susceptible fraction. The AH model involves covariates that may differ from $\mb{z}_i$, and they often contain baseline (time-fixed) and time-varying covariates. Specifically, for individual $i$, we let $\mb{w}_i=(w_{i1},\cdots,w_{ir})^\top$ to denote a vector of $r$ baseline covariates and $\mb{x}_i(t)=(x_{i1}(t),\cdots,x_{ip}(t))^\top$ to represent a vector of $p$ time-varying covariates.

The observed survival time and covariates values for individual $i$ can be represented as $(t_i^L, t_i^R, \delta_i, \delta_i^L, \delta_i^R, \delta_i^I, \mb{w}_i^\top, \tilde{\mb{x}}_i(\tilde{t}_i)^\top, \mb{z}_i^\top)$, where
$\delta_i, \delta_i^L, \delta_i^R$ and $\delta_i^I$ denote the indicators for, respectively, event, left, right and interval censoring and $\tilde{\mb{x}}_i(\tilde{t}_i)$ represents the history of $\mb{x}_i(t)$ up to time $\tilde{t}_i$. Here, $\tilde{t}_i = t_i^R$ when $t_i^R < \infty$, and $\tilde{t}_i = t_i^L$ when $t_i^R=\infty$.
Note that only when $\delta_i^R=1$ we are uncertain if $U_i=1$ or $U_i=0$, and when $\delta_i^R=0$ we have $U_i=1$.

To describe the mixture cure AH model, we adopt a mixture distribution for
$\widetilde{S}(t |\mb{w}_i, \tilde{\mb{x}}_i(t), \mb{z}_i)$, representing the overall survival function of the unconditional $T_i$ \citep{sy_estimation_2000}, namely $\widetilde{S}(t |\mb{w}_i, \tilde{\mb{x}}_i(t), \mb{z}_i) = P(T_i > t |\mb{w}_i, \tilde{\mb{x}}_i(t), \mb{z}_i)$. 
Thus, this unconditional survival function is modelled according to: 
\be \label{mixSurv}
 \widetilde{S}(t | \mb{w}_i, \tilde{\mb{x}}_i(t), \mb{z}_i) = \pi(\mb{z}_i)S(t |\mb{w}_i, \tilde{\mb{x}}_i(t), U_i=1) + 1-\pi(\mb{z}_i),
\ee
where $S(t |\mb{w}_i, \tilde{\mb{x}}_i(t), U_i=1)$ denotes the survival function corresponding to the susceptible fraction.
Similar to \cite{sy_estimation_2000}, we assume a logistic model for $\pi(\mb{z}_i)$ so that
\be \label{logit_noncure}
 \pi(\mb{z}_i) = \frac{\exp\{\mb{z}_i^\top \bsb{\gamma}\}}{1+\exp\{\mb{z}_i^\top \bsb{\gamma}\} },
\ee where $\bsb\gamma = (\gamma_1, \ldots, \gamma_q)^\top$ is the regression coefficient vector for the logistic model. The model for
$S(t|\mb{w}_i, \tilde{\mb{x}}_i(t), U_i=1)$ is defined by its corresponding hazard function model, for which we select an AH model (e.g.
\cite{LinYin94}):
\begin{equation} \label{AH model}
h(t|\mb{w}_i, \tilde{\mb{x}}_i(t), U_i=1)=h_0(t)+\mb{x}_i(t)^\top \bsb{\beta} +\mb{w}_i^\top \bsb{\alpha},
\end{equation}
where $\bsb{\beta}=(\beta_{1},\cdots,\beta_{p})^\top$ and $\bsb{\alpha}=(\alpha_{1},\cdots,\alpha_{r})^\top$,
representing regression coefficient vectors for the time-varying and the baseline covariates respectively, and $h_0(t)$
is an unspecified baseline hazard. Since both $h_0(t)$ and $h(t|\mb{w}_i, \tilde{\mb{x}}_i(t), U_i=1)$ are hazard functions,
they must be non-negative:
\be \label{constraints}
 h_{0}(t)\geq 0 \text{~~and~~} h(t|\mb{w}_i, \tilde{\mb{x}}_i(t), U_i=1)\geq 0, ~ \forall t \geq 0 \text{~and for all~$i$}.
\ee

In this paper, we assume that the values of the time-varying covariate vector $\mb{x}_i(t)$ are available
only at a finite number of time points $t_{i1}<\cdots<t_{i,n_i-1}<t_{in_i}=\tilde{t}_i$. 
We assume that there are no measurement errors among the $\mb{x}_i(t)$ values.
Therefore, each $x_{ij}(t)$ ($j = 1, \ldots, p$) is piecewise constant over intervals $(t_{i0},t_{i1}],
(t_{i1}, t_{i2}], \cdots, (t_{i,n_i-1}, t_{in_i}]$ with $t_{i0}=0$. The value of $x_{ij}(t)$ over the
interval $(t_{i, a-1}, t_{i,a}]$, where $a = 1, \ldots, n_i$, is denoted by $x_{iaj}$. Let
$\mb{x}_{ia}=(x_{ia1}, \cdots, x_{iap})^\top$. Define matrix $\mb{X}_i = (\mb{x}_{i1}, \ldots,
\mb{x}_{in_i})^\top$, which has the dimension of $n_i\times p$, and matrix $\mb{X} = (\mb{X}_1^\top, \ldots,
\mb{X}_n^\top)^\top$, which has the dimension of $(\sum_{i=1}^n n_i)\times p$. Under this piecewise constant
assumption, the hazard expression in (\ref{AH model}), when $t \in (t_{i,a-1}, t_{ia}]$, becomes
\begin{equation} \label{AH model PC}
  h(t|\mb{w}_i, \tilde{\mb{x}}_i(t), U_i=1)=h_0(t)+\mb{x}_{ia}^\top \bsb{\beta} +\mb{w}_i^\top \bsb{\alpha},
\end{equation}
The corresponding cumulative hazard function is then
\begin{equation}
  H(t|\mb{w}_i, \tilde{\mb{x}}_i(t), U_i=1) =H_0(t)+\mb{X}_{ia}(t)^\top \bsb{\beta}+\mb{w}_i^\top \bsb{\alpha}\,t,
\end{equation}
where $H_0(t) = \int_0^t h_0(s)ds$ is the cumulative baseline hazard, and
$\mb{X}_{ia}(t)=\sum_{b=1}^{a-1}\mb{x}_{ib}(t_{ib}-t_{i,b-1})+\mb{x}_{ia}(t-t_{i,a-1})$ when
$t \in (t_{i,a-1}, t_{ia}]$. Note $\mb{X}_{ia}$ is a $p$-vector. When there is no confusion, we will denote the hazard, cumulative hazard, and
survival function of subject $i$, if $i$ is in the susceptible fraction, by $h_i(t)$, $H_i(t)$, and $S_i(t)$,
respectively.

Estimation of $h_0(t)$, an infinite dimensional parameter, from a finite number of observations is an
ill-posed problem. A common approach to handle this issue is to approximate $h_0(t) \geq 0$ using basis
functions $\psi_u(t) \geq 0$ $(u = 1, \ldots, m)$, such that
\begin{equation} \label{happ}
 h_{0}(t)=\sum_{u=1}^{m}\theta_{u}\psi_u(t),
\end{equation}
where the number of basis functions, $m$, can vary with the sample size $n$. This is also known as the
method-of-sieves to estimate a nonparametric function; see \cite{Grenander81} and \cite{GeHw82}. Widely used
non-negative basis functions include M-splines, Gaussian density functions or indicator functions
\citep{MaCoHeIa21}. Now, the constraint $h_0(t) \geq 0$ can be replaced with an easier requirement that
$\theta_u\geq 0$ for all $u$. Basis functions are defined using a set of  preselected knots.

Let $\bsb{\theta}=[\theta_{1},\cdots,\theta_{m}]^\top$ and $\bsb\psi(t) = (\psi_1(t), \ldots, \psi_m(t))^\top$. The
constraints we wish to impose are: $\theta_1\geq 0, \ldots, \theta_m\geq 0$ and $ \bsb\psi(t)^\top \bsb\theta
+ \mb{x}_i(t)^\top \bsb\beta+ \mb{w}_i^\top \bsb\alpha \geq 0 \text{~for all t and~} i$, where the latter set of
constraints are difficult to impose as they involve all $t$. 
To make these constraints feasible, we impose 
them only at the $\mb{x}_i(t)$
measurement time points $t_{ia}$. Therefore, we now have 
the following more manageable set of $\sum_{i=1}^n n_i$ constraints:
\be
  \bsb\psi(t_{ia})^\top\bsb\theta + \mb{x}_i(t_{ia})^\top \bsb\beta + \mb{w}_i^\top \bsb{\alpha}\geq 0
\ee
where $a = 1, \ldots, n_i$ and $i = 1, \ldots, n$. Similar to \cite{LiMa19}, we will discuss a primal-dual
algorithm in Section \ref{sec:algo} to compute the constrained optimal solutions.

Let $\bsb{\eta}=[\bsb{\theta}^\top, \bsb{\beta}^\top, \bsb\alpha^\top, \bsb\gamma^\top]^\top$.
Under the assumption that the partly interval-censored survival data are independent, the log-likelihood is given by:
\begin{align} \label{lliki}
 \ell(\bsb{\eta}) = & \sum_{i=1}^n \Big\{\delta_i\{\log\pi_i + \log h_i(t_i) - H_i(t_i)\} +
 \delta_i^R \log(1-\pi_i+\pi_i S_i(t_i^L))  \nonumber\\
  & + \delta_i^L\{\log\pi_i+\log(1- S_i(t_i^R))\}+ \delta_i^I\{\log\pi_i + \log(S_i(t_i^L)-S_i(t_i^R))\}\Big\},
\end{align}
where $t_i$ is the event time when $t_i^L = t_i^R=t_i$. We estimate $\bsb\gamma$, $\vect{\beta}$, $\bsb\alpha$ and $\vect{\theta}$ by maximizing the
following penalized log-likelihood
\begin{equation} \label{plik}
 \Phi(\bsb{\eta})=\ell(\bsb{\eta})-\omega J(\bsb\theta),
\end{equation}
where $J(\vect{\theta})$ is a penalty function for obtaining a smooth estimate of $h_0(t)$ and $\omega\geq0$
is a smoothing parameter. In this paper we adopt the roughness penalty (e.g. \citet{GreSil94}) given by
$J(\vect{\theta})=\int h_{0}''(t)^2dt=\vect{\theta}^\top\mb{R}\vect{\theta}$, where $\mb{R}$ is an $m\times
m$ matrix with the $(u,v)$-th element $r_{uv}=\int\psi''_{u}(t)\psi''_{v}(t)dt$. We comment that a penalty
function is also helpful to make the estimates less dependent on the location and number of knots, and thus
stabilizes the estimation process with less numerical errors; see for example \cite{MaCoHeIa21} and
\cite{RuWaCa03}.

\section{
Constrained optimal solution} \label{sec:algo}

\subsection{Computation} \label{sec:algoInteriorPoint}
The vector $\bsb\eta$ contains $v = m+p+r+q$ elements, and it is necessary to impose $w = m+\sum_i n_i$ inequality constraints on it, as indicated by (\ref{fs}). 
The constrained optimization problem specified above can be formulated as
\begin{equation}
\widehat{\boldsymbol{\eta}}=\operatorname*{argmax}_{\boldsymbol{\eta}\in\mathcal{F}}\left\{\Phi(\boldsymbol{\eta})\right\},
\label{co}
\end{equation}
where $\Phi(\bsb{\eta})$ is defined in (\ref{plik}) and $\mathcal{F}$ is a feasible set for the constraints defined by
\begin{equation}
\mathcal{F}=\left\{\boldsymbol{\eta}\,|\,f_{b}(\boldsymbol{\eta})=-\mb{M}_{b}\boldsymbol{\eta}\leq 0, ~b=1,\ldots,w\right\},
\label{fs}
\end{equation}
where the details of $\mb{M}_b$ can be found in the Supplementary Materials of this paper. Each $\mb{M}_b$ is a row vector of length $v$. Let matrix $\mb{M}=(\mb{M}_1^\top, \cdots, \mb{M}_w^\top)^\top$ (its dimension is $w \times v$) and $f(\bsb\eta) = -\mb{M}\bsb\eta$.
We use the primal-dual interior-point method to compute the constrained optimal solution for our problem. This method is briefly explained below; its details can be found in
the Supplementary Materials.

Let $\lambda_b$ $(\geq 0)$, $b=1, \ldots, w$, be the Lagrange multipliers for the inequality constraints and $\vect{\lambda}$ be the vector for all $\lambda_b$.
We employ
slack variables $s_b\geq 0$ such that these inequality constraints become equality constraints. Let $\vect{s}$ be the vector for all $s_b$. The constrained optimization (\ref{co}) can be achieved by solving the following modified Karush-Kuhn-Tucker (KKT) conditions (e.g. Chapter 8 of  \cite{sun2006optimization}):
\begin{align}
&\nabla\Phi(\boldsymbol{\eta})+\mb{M}^\top\boldsymbol{\lambda}
=\boldsymbol{0}_{v \times 1},
\label{k1} \\
&f(\boldsymbol{\eta})+\boldsymbol{s}=\boldsymbol{0}_{w \times 1},
\label{k2}\\
&\vect{\lambda} * \vect{s}-\kappa\vect{1}_{w \times 1}=\boldsymbol{0}_{w \times 1},
\label{k3} \\
&\vect{\lambda}\geq0, \vect{s}\geq0,
\kappa\geq 0,
\label{k4}
\end{align}
where the operator `$*$' represents elementwise multiplications, $\nabla$ denotes the derivative, $\bsb{\lambda}\geq 0$ and $\bsb{s}\geq 0$ mean each element of these vectors are non-negative, $\bsb{1}$ is a vector of 1's and $\bsb{0}$ is a vector of 0's. Equation (\ref{k3}) reflects the modification to the standard KKT conditions which require $\bsb{\lambda} * \vect{s}=\vect{0}$, known as the slackness condition. Equation (\ref{k3}) is a perturbed version of this condition,
and it helps to achieve more stable computations of the solution.
Note that as $\kappa$ goes to zero, $\vect{\lambda} * \vect{s}$ also goes to zero, and thus the slackness condition will be satisfied.

Following \cite{LiMa19}, we propose to  solve the system (\ref{k1}) -- (\ref{k4}) using a primal-dual interior-point algorithm. It solves simultaneously the primal vector $\boldsymbol{\eta}$, and the dual vectors $\boldsymbol{\lambda}$ and $\boldsymbol{s}$. This algorithm assumes $\kappa$ takes the form $\kappa=\xi\mu$ where $\xi\in [0,1]$ and $\mu~(\geq 0)$ is defined by
$
\mu=\boldsymbol{\lambda}^\top \boldsymbol{s}/w,
$
measuring the average value of $\lambda_b s_b$. The value of $\mu$ will be gradually shrank towards 0 during the iterations.
Parameter $\xi$ is called a centering parameter and $\mu$ is called a duality measure.
Rather than estimating $\kappa$ directly, the algorithm estimates $\xi$ and $\mu$. 
This type of algorithm has been studied in depth in \cite{sun2006optimization}. \cite{Ghosh01} and \cite{LiMa19} applied similar algorithms in studying the AH model with, respectively, current status and partly interval censoring data.

Let $\vect{\eta}^{(k)}$, $\vect{\lambda}^{(k)}$ and $\vect{s}^{(k)}$ be the values for $\bsb\eta$, $\bsb\lambda$ and $\bsb s$ at iteration $k$. The corresponding duality measure is
$\mu^{(k)}=(\boldsymbol{\lambda}^{(k)})^\top\boldsymbol{s}^{(k)}/w$. At the $(k+1)$th iteration, we first calculate the
direction $(d\vect{\eta}, d\vect{\lambda}, d\vect{s})$ defined by the
Newton algorithm, which is equivalent to solve the linear system (G2) in the Supplementary Materials.
Then, the updates for $\bsb\eta$, $\bsb\lambda$ and $\bsb s$ are obtained according to
\begin{equation}
(\boldsymbol{\eta}^{(k+1)},\boldsymbol{\lambda}^{(k+1)},\boldsymbol{s}^{(k+1)})=
(\boldsymbol{\eta}^{(k)},\boldsymbol{\lambda}^{(k)},\boldsymbol{s}^{(k)})+
\alpha_{k}(d\boldsymbol{\eta}^{(k)},d\boldsymbol{\lambda}^{(k)},d\boldsymbol{s}^{(k)}),
\label{ups}
\end{equation}
where $\alpha_{k}$ is a step length chosen to be the first element in the sequence
$
\{1,\epsilon, \epsilon^2, \epsilon^3, \ldots\},
$
($\epsilon\in (0,1)$, e.g. $\epsilon=0.6$) meeting the following two conditions: $(\vect{\eta}^{(k+1)},\vect{\lambda}^{(k+1)},\vect{s}^{(k+1)})\in\mathcal{N}(\mu^{(k)})$
and also
$
\mu^{(k+1)}\leq(1-0.01\alpha_k)\mu^{(k)}.
$
Here, $\mathcal{N}(\mu)$ represents a closed ball for $\bsb\eta$, $\bsb\lambda$ and $\bsb s$, with radius $\zeta\mu$ and centered at the solution, where $\zeta>0$ is a small fixed quantity such as $10^{-1}$. The detailed definition of $\mathcal{N}(\mu)$ can be found in (G1) of the Supplementary Materials.

By applying Theorem 6.1 of \cite{Wright97}, a global convergence result of this primal-dual interior-point algorithm can be obtained. Essentially, the sequence ${\boldsymbol{\eta}^{(k)},\boldsymbol{\lambda}^{(k)},\boldsymbol{s}^{(k)}}$ generated by the primal-dual interior-point algorithm converges to a solution that satisfies conditions (\ref{k1}) -- (\ref{k4}).

\subsection{Smoothing parameter estimation} \label{sec:smpa}
For the MPL method, it is important to find
a suitable value for the smoothing parameter $\omega$ that will provide a good balance between
fit to the data and smoothness of the MPL baseline hazard estimates.
Similar to \cite{MaCoHeIa21}, We will discuss a marginal likelihood method for smoothing parameter selection where the penalty is in a quadratic form of
$J(\vect{\theta})=\vect{\theta}^{\top}\vect{R}\vect{\theta}$.
This quadratic penalty of $\bsb\theta$ can be conceived as
a log-normal density where $\bsb{\theta} \sim N(\mb{0}_{m\times 1},\sigma^2\mb{R}^{-1})$, where $\sigma^2=1/(2\omega)$.

If treating this
distribution of $\bsb\theta$ as a prior distribution, the log-posterior becomes
\[
\ell_{p}(\bsb{\beta}, \bsb\alpha, \bsb\gamma, \bsb{\theta})=-\frac{m}{2}\log\sigma^2+\ell(\bsb{\beta}, \bsb\alpha, \bsb\gamma, \bsb{\theta})-\frac{1}{2\sigma^2}\bsb{\theta}^\top\vect{R}\bsb{\theta},
\]
where terms independent of $\bsb{\theta}, \bsb{\beta}, \bsb\alpha, \bsb\gamma$ and $\sigma^2$ are omitted.
Note that when $\sigma^2$ is given, $\ell_{p}(\bsb{\beta}, \bsb\alpha, \bsb\gamma, \bsb{\theta})$ is equivalent to the penalized log-likelihood given in (\ref{plik}).
The log-marginal likelihood for $\sigma^2$ (after integrating out $\bsb{\beta}, \bsb\alpha, \bsb\gamma \text{~and~} \bsb{\theta}$) is
\[
\ell_{m}(\sigma^2)=-\frac{m}{2}\log\sigma^2+\log\int\exp\{\ell(\bsb{\beta}, \bsb\alpha, \bsb\gamma, \bsb{\theta})-\frac{1}{2\sigma^2}\bsb{\theta}^\top\mb{R}\bsb{\theta}\} d\bsb{\beta} d\bsb\alpha d\bsb\gamma d\bsb \theta.
\]
Letting $\widehat{\bsb{\beta}}$, $\widehat{\bsb{\alpha}}$, $\widehat{\bsb{\gamma}}$ and $\widehat{\bsb{\theta}}$ be the values maximizing
$\ell_p(\bsb{\beta}, \bsb\alpha, \bsb\gamma, \bsb{\theta})$, then, after applying the Laplace's approximation, we have
\begin{equation}
\ell_{m}(\sigma^2)\approx-\frac{m}{2}\log\sigma^2+\ell(\widehat{\bsb{\beta}}, \widehat{\bsb{\alpha}}, \widehat{\bsb{\gamma}}, \widehat{\bsb{\theta}})-\frac{1}{2\sigma^2}\widehat{\vect{\theta}}^\top\vect{R}\widehat{\vect{\theta}}-\frac{1}{2}\log|\widehat{\vect{G}}+\vect{Q}(\sigma^2)|,
\label{5(3)}
\end{equation}
where $\widehat{\vect{G}}$ is the negative Hessian from $\ell(\bsb{\beta}, \bsb{\alpha}, \bsb{\gamma}, \bsb{\theta})$, evaluated at $\widehat{\vect{\beta}}$,
$\widehat{\bsb{\alpha}}, \widehat{\bsb{\gamma}}$ and $\widehat{\vect{\theta}}$, and
\[
\vect{Q}(\sigma^2)=
\begin{pmatrix}
\mb{0}_{p\times p} & \mb{0}_{p\times r} & \mb{0}_{p\times q} & \mb{0}_{p\times m}\\
\mb{0}_{r\times p} & \mb{0}_{r\times r} & \mb{0}_{r\times q} & \mb{0}_{r\times m}\\
\mb{0}_{q\times p} & \mb{0}_{q\times r} & \mb{0}_{q\times q} & \mb{0}_{q\times m}\\
\mb{0}_{m\times p} & \mb{0}_{m\times r} & \mb{0}_{m\times q} & \frac{1}{\sigma^2}\mb{R}
\end{pmatrix},
\]
where $\mb{0}$ denotes a matrix of zeros with dimension specified in the subscript.

The solution for $\sigma^2$ that maximizes equation (\ref{5(3)}), denoted by $\widehat{\sigma}^2$, can be verified to satisfy:
\begin{equation}
\widehat{\sigma}^2=\frac{\widehat{\vect{\theta}}^\top\vect{R}\widehat{\vect{\theta}}}{m-\nu},
\label{5(4)}
\end{equation}
where $\nu$ is given by:
\[
 \nu=\text{tr}\{\big(\widehat{\vect{G}}+\vect{Q}(\widehat{\sigma}^2)\big)^{-1}
 \vect{Q}(\widehat{\sigma}^2)\}.
\]
Note that $m - \nu$ is equivalent to the model degrees of freedom.
Because $\bsb{\beta}, \bsb\alpha, \bsb\gamma$, and $\bsb{\theta}$ depend on $\sigma^2$, the expression (\ref{5(4)}) suggests an iterative procedure.
Specifically, with $\sigma^2$ fixed at its current estimate,
the corresponding MPL estimates of $\bsb{\beta}, \bsb\alpha, \bsb\gamma$ and $\bsb{\theta}$ are obtained. Then, $\sigma^2$
is updated using formula (\ref{5(4)}), where $\widehat{\bsb{\beta}}$, $\widehat{\bsb{\alpha}}$, $\widehat{\bsb{\gamma}}$, $\widehat{\bsb{\theta}}$,
and $\widehat{\sigma}^2$ on the right-hand side are replaced by their most current estimates.
These iterations continue until the degree of freedom stabilizes,
i.e. difference between consecutive degrees of freedom is less than 1.

\section{Asymptotic properties of the MPL estimators
} \label{sec:asym}
Development of asymptotic properties of the mixture cure AH model
allows for large sample inferences to be conducted without computing intensive methods
such as bootstrapping. Following \cite{LiMa19}, it is possible to demonstrate asymptotic
consistency for the MPL estimates of regression coefficients $\bsb\gamma$, $\bsb\beta$
and $\bsb\alpha$, and also for the estimate of the baseline hazard $h_0(t)$. We use
$\bsb\gamma_0$, $\bsb\beta_0$, $\bsb\alpha_0$ and $h_{00}(t)$ to represent the true
parameters. In Theorem \ref{the1}, $a = \min\{t_i\}$, $b = \max\{t_i\}$ and $\rho_n =
\omega /n$.

\begin{theorem} \label{the1}
Suppose that the Assumptions A1-A3 in the Supplementary Materials hold. Assume that $h_0(t)$ is bounded and has up to $r \geq 1$ derivatives over the interval [a, b].
Assume $m=n^\nu$ where $0<\nu<1$, and $\rho_n \to 0$ when $n \to \infty$. Then,
as $n\to\infty$,
\begin{enumerate}
\item $\|\widehat{\bsb{\gamma}}-\bsb{\gamma}_0\|_2 \to 0$ (a.s.),
\item $\|\widehat{\bsb{\beta}}-\bsb{\beta}_0\|_2 \to 0$ (a.s.),
\item $\|\widehat{\bsb{\alpha}}-\bsb{\alpha}_0\|_2 \to 0$ (a.s.), and
\item $\displaystyle \sup_{t \in [a, b]}|\widehat{h}_{0}(t)-h_{0}^0(t)|\to 0$ (a.s.).
\end{enumerate}
\end{theorem}
\noindent Proof: See the Supplementary Materials Section S1.

Apart from the above asymptotic consistency results, it is also desirable to develop asymptotic normality results for all parameters $\bsb\gamma$, $\bsb\beta$ and $\bsb\alpha$ and
$\bsb\theta$ as this will facilitate inferences
to be made not only on regression coefficients but also on other quantities of interest, such as survival probabilities. In
order to develop these results, however, it is necessary to restrict $m$ to be a finite number as similar to \cite{YuRup02} and \cite{MaCoHeIa21}.
It is important to note that this $m$ depends on the given sample size $n$.
A general practical guide for $m$ we recommend is
$m = \sqrt[3]{n}$.

Recall $\bsb{\eta}
$ is the vector for all the parameters, and let
$\bsb{\eta}_0=(\bsb{\theta}_0^\top, \bsb{\beta}_0^\top, \bsb\alpha_0^\top, \bsb\gamma_0^\top)^\top$ be the true value of $\bsb{\eta}$.
For the asymptotic normality result, we closely follow  \cite{moore08} to
address the active constraints in the MPL estimates.
Recall that the constraints are $\mb{M}\vect{\eta}\geq 0$, where matrix $\mb{M}$ is defined in Section \ref{sec:mple}.
From the KKT condition in equation (\ref{k1}) we have
\begin{equation}
 \frac{\partial
\Phi}{\partial \bsb{\eta}} + \mb{M}^\top \vect{\lambda} = 0,
\end{equation}
where $\bsb{\lambda}$ is the vector for all Lagrange multipliers $\lambda_b$.
Here, $\lambda_b > 0$
if the corresponding constraint is active and $\lambda_b=0$ for otherwise.
We divide $\bsb{\lambda}$ and $\mb{M}$ into sub-matrices according to active and non-active constraints. The portion of $\bsb\lambda$ for
active constraints is denoted by $\bsb{\lambda}_A$, and
the corresponding portion of $\mb{M}$ denoted by $\mb{M}_A$.
Since $\mb{M}^\top \bsb{\lambda} = \mb{M}_A^\top \bsb{\lambda}_A $, we will only focus on the $\mb{M}_A$ matrix in the following discussions.
Assume there are $r \geq 1$ active constraints, then the dimension of $\mb{M}_A$ is $r \times v$. Usually, $r$ is a small
number so we can safely assume $r < v$. Under this assumption,
the null space of $\mb{M}_A$ is non-empty, which means there are non-zero vectors $\mb{u}$ 
(of length $v$) satisfying $\mb{M}_A \mb{u}=0$.
Let $q$ denote the dimension of this null space. Let $\mb{U}_{v\times q}$ be a matrix whose columns form orthonormal bases
of the null space of $\mb{M}_A$. Matrix $\mb{U}$ satisfies
\begin{equation}
 \mb{M}_A \mb{U} = \mb{0}_{r\times q} \text{~and~} 
\mb{U}^\top \mb{U}=\mb{I}_{q\times q}.
\end{equation}
Now, we are ready to state the following asymptotic normality theorem for the MPL estimate $\widehat{\bsb{\eta}}$ where $\bsb\eta_0$ denotes the true $\bsb\eta$ with a fixed $m$.

\begin{theorem} \label{the2}
Suppose Assumptions A4-A7 in the Supplementary Materials hold. Recall $\rho = \omega/n$ and assume $\rho \to 0$ when $n \to \infty$.
Let $\mb{G}(\vect{\eta}) = -\lim_{n\to\infty} n^{-1} E_{\bsb{\eta}_0} \partial^2 \ell/\partial \bsb{\eta}\partial \bsb{\eta}^\top$.
Assume there are $r \neq 0$ active constraints and define matrix $\mb{U}$ as above.
Then, when $n \to \infty$,
\begin{enumerate}
\item The constrained MPL estimate $\widehat{\bsb{\eta}}$ is consistent for $\bsb{\eta}_0$, and
\item $n^{1/2} (\widehat{\bsb{\eta}} -\bsb{\eta}_0 )$ converges in distribution to a multivariate normal
$N(\vect{0}_{v\times 1}, \mb{W}(\bsb{\eta}_0))$, where the covariance matrix
$\mb{W}(\bsb{\eta}_0) =\mb{U}(\mb{U}^\top\mb{G}(\bsb{\eta}_0)\mb{U})^{-1} \mb{U}^\top$.
\end{enumerate}
\end{theorem}
\noindent Proof: See the Supplementary Materials section S1.

Since $\bsb{\eta}_0$ is generally unavailable, we replace it by $\widehat{\vect{\eta}}$ due to the strong
consistent result. The expected information matrix $\mb{G}(\bsb{\eta})$ can also be difficult to obtain so we usually replace it with the negative Hessian matrix of the
log-likelihood function $\ell(\bsb\eta)$.

The inferences are typically conducted for finite values of $n$, and hence, the penalty
term needs to be included in the asymptotic covariance matrix formula provided in
Theorem \ref{the2}. To modify the results in Theorem \ref{the2} for a large but finite
$n$, we use an approximate distribution for $\widehat{\bsb{\eta}} - \bsb{\eta}_0$, which
is a multivariate normal distribution with a zero mean and a covariance matrix
$\mb{V}(\widehat{\bsb{\eta}})$, given by:
\be \mb{V}(\widehat{\bsb{\eta}})=
\mb{U}(\mb{U}^\top(\partial^2 \ell(\widehat{\bsb{\eta}})/\partial \bsb{\eta}
\partial\bsb{\eta}^\top+ \omega \partial^2 J(\widehat{\bsb{\eta}})/\partial
\bsb{\eta}\partial\bsb{\eta}^\top)\mb{U})^{-1}\mb{U}^\top. \label{asyvar}
\ee
The simulation results reported in Section \ref{sec:resu} demonstrate this asymptotic covariance matrix
is generally accurate even for small to moderate sample sizes.

\section{Results} \label{sec:resu}
In this section we report a simulation study that evaluates our penalized likelihood method for fitting the mixture cure AH model. The main aim of this simulation is to demonstrate that our method
is capable of producing accurate results on regression coefficient and baseline hazard, measured by bias, mean squared error (MSE) and coverage probability. Accuracy of the asymptotic standard
deviation of the MPL estimates of coefficient $\bsb\beta$ is also assessed through comparisons of the average asymptotic standard deviation with the Monte-Carlo standard deviation. We will implement our method to a thin melanoma data set where the event-of-interest is melanoma recurrence after treatment. All results in this section were obtained using the statistical software R, and all the simulation codes are available at GitHub: \texttt{https://github.com/ActuaryJin}.

\subsection{Simulation results}

A simulation study was conducted (i) to investigate the effects of cure rate, censoring proportion and sample size on our proposed MPL estimates of $\vect{\beta}$, $\alpha$, $\vect{\gamma}$ and $h_0(t)$, and (ii) to evaluate the accuracy of the asymptotic standard deviations given in (\ref{asyvar}), achieved by comparing the average asymptotic standard deviation with the Monte Carlo standard deviation.

For the logistic regression model (\ref{logit_noncure}) in our simulation, we generated a binary covariate using a Bernoulli distribution $z_{i1}\sim\text{Bernoulli}(0.5)$, and
a continuous covariate using a uniform distribution $z_{i2}\sim\text{Uniform}(d_1,d_2)$. The values for $d_1$ and $d_2$ were used to control the size of the non-cured fraction. The logistic regression coefficients were set to $\gamma_{1}=-0.2$ and $\gamma_{2}=0.5$, and the values of $(d_1, d_2)=(3, 3.5)$ and $(d_1, d_2)=(1, 1.2)$ produced non-cured fraction proportions of 80\% and 60\%, respectively. Prior to generating observed survival times, an indicator value $u_i$ was computed as follows: firstly, generated a $\zeta_i \sim \text{Uniform}(0, 1)$ and computed $\pi(\mb{z}_i)$ 
from $\mb{z}_i$ and the logistic model. Then, $u_i$ was set to 1 if $\zeta_i\leq \pi(\mb{z}_i)$, indicating that individual $i$ was susceptible to the event-of-interest, and $u_i$ was set to 0 if $\zeta_i> \pi(\mb{z}_i)$, indicating that individual $i$ was cured.

For individuals in the cured group, their observed times were merely right censoring times, drawn from the $\text{Uniform}(0, 2.5)$ distribution.
For individuals in the non-cured group, their event times were simulated using
model (\ref{AH model})
where there were three covariates: two time-fixed and one time-varying. The two time-fixed covariates were: $w_{i1}=z_{i1}$, where $z_{i1}$ was the first covariate for the logistic model above, and $w_{i2}\sim\text{Uniform}(1,2)$. The time-varying covariate $x_{i1}(t)$ took value 0  from $t = 0$ until a random selected time point $t_i^*$ (i.e. $x_{i1}(t) = 0$ for $t \leq t_i^*$) and then it switched to value 1 (i.e
$x_{i1}(t) = 1$ for $t>t_i^*$).
We drew $t_i^*$ from the $\text{Uniform}(0.5,2.5)$ distribution.
For the additive hazard function in the simulation, we adopted the baseline hazard $h_0(t)=3t^2$, the coefficients for the time-fixed covariates were $\alpha_1 = -0.2$ and $\alpha_2 = 0.3$ and the coefficient for the time-varying coefficient was $\beta_1 = 0.5$.
Note that this $h_i(t)$ is guaranteed to be non-negative for all $i$ and $t$. The simulated observed survival times in this non-cured
group contained event times, left censoring times, finite interval censoring times and right censoring times. For individual $i$ in this group, we first generated $\tau_i \sim \text{Uniform}(0, 1)$, then
the event time $t_i$ was obtained by solving $H_i(t) + \log \tau_i = 0$, where
the cumulative hazard $H_i(t)$ is:
$
H_i(t) =t^3+(\alpha_{1}w_{i1}+\alpha_{2}w_{i2})t
$
if $t\leq t_{i}^*$,
and
$
H_i(t) =t^3+(\alpha_{1}w_{i1}+\alpha_{2}w_{i2})t + \beta_{1}(t-t_{i}^*)
$
if $t> t_{i}^*$.
Due to censoring, $t_i$ might not be given exactly.
We let $(t_i^L, t_i^R]$ be a general censoring interval associated with $t_i$, where $t_i^L$ and $t_i^R$ were generated as the following.
Let $\pi_c$ be the censoring proportion (including left, right and interval censoring) for this non-cured group. Firstly, two independent time points were computed according to: $L_i\sim\text{Exp}(3)$ and $R_i=L_i+\text{Uniform}(0,1)$.
We then compared $\pi_c$ with a standard uniform random number, denoted by $U_{i}^C$, to decide if $t_i$ was fully observed. If $U_{i}^C\geq \pi_c$, $t_i$ was fully observed so that $t_{i}^L=t_{i}^R=t_i$. If $U_{i}^C<\pi_c$, $t_i$ was censored and there were three cases: if $t_i\leq L_{i}$, it was left-censored at $L_i$ and we set $t_i^L=0$ and $t_{i}^R=L_{i}$; if $L_{i}<t_i\leq R_{i}$, it was interval-censored with $t_{i}^L=L_{i}$ and $t_{i}^R=R_{i}$; if $t_i>R_{i}$, it was right-censored at $R_i$ so that $t_{i}^L=R_{i}$ and $t_{i}^R=+\infty$.
In the simulation, we set $n=200$, $500$ and $1000$ representing small, intermediate and large sample sizes, respectively.

The simulation results for the regression coefficients contain
the following quantities: the absolute bias (ABIAS) given by the average of the estimates minus the true parameter value, the Monte Carlo standard deviation (MCSD), the average asymptotic standard deviation (AASD), the mean squared error (MSE), and the asymptotic 95\% coverage probability (CP).
The simulation results we report for the estimates of $h_0(t)$ includes: the ABIAS, MCSD, AASD, MSE and the 95\% coverage probabilities for $h_0(t)$ estimates at the 25th ($t_1$), 50th ($t_2$) and 75th ($t_3$) percentile of all the simulated $t_i$'s. Also, we report the average integrated squared error (AISE), given by the average of the integrated squared error (ISE):
\[
D(h_0,\widehat{h}_{0})=\int_{t\in [0, t_0]}[h_0(t)-\widehat{h}_{0}(t)]^2dt,
\]
where $t_0$ corresponds to the 90th percentile of the observed survival times.

We adopted indicator basis functions to approximate $h_0(t)$, leading to a piecewise constant approximation to $h_0(t)$. More specifically, the interval $[0, b]$, where
$b = \max\{t_i^R\}$ with
$t_i^R \neq \infty$, was divided into $m$ sub-intervals (called bins) and the approximate $h_0(t)$ took a constant value over each bin. In this context, $h_0(t)$ was represented by $\bsb{\theta}=(\theta_{1},\cdots,\theta_{m})^\top$. Bins were selected in such a way that each bin contained an approximately equal number of ``observations'' denoted by $n_o$. Here, observations included observed event times, left-censoring times and interval censoring times (each interval contributed two time points).
We set $n_o=2$ for $n=200$, $n_o=3$ for $n=500$, and $n_o=4$ for $n=1000$.
Our experience indicates that the regression coefficients estimates are generally not very sensitive to $n_o$ as long as it is not too large and the smoothing parameter is chosen appropriately.
Note that there could be other ways of defining
observations. For example, they could include observed event times, mid-points of left censoring intervals and mid-points of finite interval censoring intervals. We had also performed some simulations corresponding to this definition of observations, and the results are available in the Supplementary Materials.
There are no obvious differences in the regression coefficient estimates
from these two definitions of observations. The smoothing parameter was selected automatically as described in Section \ref{sec:smpa}. Since the approximate $h_0(t)$ was piecewise constant, we used a second order difference quadratic penalty function
$ 
J(\vect{\theta})=\sum_{j=2}^{m-1}(\theta_{j-1}-2\theta_{j}+\theta_{j+1})^2
$ 
in the simulation.
The convergence criterion for the proposed primal-dual interior-point algorithm
was: if $\mu^{(k)}<10^{-8}$ then iterations were terminated.

\begin{table}[th]
{\small
\centering
\begin{tabular}{l@{\hspace{-0.01in}}rr@{\hspace{0.001in}}rr@{\hspace{0.001in}}rr@{\hspace{0.001in}}rr@{\hspace{0.001in}}r@{\hspace{0.001in}}r}
\hline
\hline
& &\multicolumn{2}{c}{$n=200$}&\multicolumn{2}{c}{$n=500$}&\multicolumn{2}{c}{$n=1000$}\\
$\pi(\vect{z})$       &          &  60\%  &  80\%  & 60\%  & 80\% & 60\% &80\% \\
$n_{o}         $         &          &    2    &   2       &  3        &  3      &  4      &  4  \\

$\pi_R$       &          &   53\% &  37\%  &  53\% &  37\% &  53\%  &  37\%  \\
\hline
$\alpha_{1}=-0.2$
              & ABIAS       & 0.0277& 0.0641&0.0305 &0.0101&0.0110&0.0092\\
              & MCSD        & 0.2489 & 0.2177 &0.1491&0.1247&0.1154&0.0922\\
              & AASD       & 0.2745 & 0.2244 &0.1581&0.1287&0.1089&0.0893\\
              & MSE        &(0.0627)&(0.0515)&(0.0232)&(0.0156)&(0.0134)&(0.0086)\\
              &CP           & 0.96      &   0.96     &  0.97    &   0.96   &  0.94    &     0.94         \\
\\
$\alpha_{2}=0.3$
              & ABIAS       &0.1501 & 0.1283 & 0.1089&0.0860&0.0790&0.0510\\
              & MCSD        & 0.2234 & 0.1837 & 0.1595&0.1295&0.1285&0.1027\\
              & AASD       & 0.2761 & 0.2802 & 0.1691&0.1592&0.1814&0.1472\\
              & MSE        &(0.0724)&(0.0502)&(0.0373)&(0.0242)&(0.0227)&(0.0131)\\
              &CP           & 0.99      &    0.98      &   0.98  &   0.97   & 0.96     &  0.97            \\
\\
$\beta_{1}=0.5$
              & ABIAS       & 0.1769 &0.1362 &0.0986&0.0707&0.0839&0.0572\\
              & MCSD        & 0.3901 & 0.2708 & 0.2067&0.1840&0.1583&0.1128\\
              & AASD       & 0.3810 & 0.2706 & 0.2203&0.1837&0.1601&0.1171\\
              & MSE        &(0.1835)&(0.0919)&(0.0524)&(0.0389)&(0.0321)&(0.0160)\\
              &CP           & 0.99      &   0.96         &   0.97  &    0.93  &  0.93    &   0.95           \\
\\
$\gamma_{1}=-0.2$
              & ABIAS       & 0.0971 &0.0537 &0.0233&0.0017&0.0039&0.0020\\
              & MCSD        & 0.3964 & 0.3895 &0.2436 &0.2441& 0.1737 &0.1129\\
              & AASD       &0.3899 & 0.3868 & 0.2377&0.2220&0.1663 &0.1220\\
              & MSE        &(0.1574)&(0.1546)&(0.0599)&(0.0596)&(0.0302)&(0.0128)\\
              &CP           & 0.96      &   0.98     &   0.96   &    0.94  &  0.96   &    0.96          \\
\\
$\gamma_{2}=0.5$
              & ABIAS       & 0.0630  &0.0180 &0.0142 &0.0376 &0.0125 &0.0095 \\
              & MCSD        & 0.2356 &0.1411  & 0.1521&0.0770 &0.1039&0.0460\\
              & AASD       & 0.2537 &0.1244 & 0.1536&0.0727 &0.1081&0.0496\\
              & MSE        &(0.0595)&(0.0202)&(0.0233)&(0.0073)&(0.0109)&(0.0022)\\
              &CP           & 0.97      &   0.97      &  0.93    &    0.93 &     0.94   &   0.96           \\
\hline
\end{tabular}
\caption{ABIAS, MCSD, AASD and MSE of $\bsb\alpha$, $\vect{\beta}$ and $\vect{\gamma}$ for samples with $\pi_c=70\%$.
}
\label{tab1}
}
\end{table}

\begin{table}[h]
{\small
\centering
\begin{tabular}{l@{\hspace{-0.01in}}rr@{\hspace{0.001in}}rr@{\hspace{0.001in}}rr@{\hspace{0.001in}}rr@{\hspace{0.001in}}r@{\hspace{0.001in}}r}
\hline
\hline
& &\multicolumn{2}{c}{$n=200$}&\multicolumn{2}{c}{$n=500$}&\multicolumn{2}{c}{$n=1000$}\\
$\pi(\vect{z})$       &          &  60\%  &  80\%  & 60\%  & 80\% & 60\% &80\% \\
$n_{o}         $         &          &    2    &   2       &  3        &  3     &  4      &  4  \\

$\pi_R$       &          &   47\% &  29\%  &  47\% &  29\% &  47\%  &  29\%  \\
\hline
$\alpha_{1}=-0.2$
              & ABIAS       & 0.0160& 0.0144&0.0210&0.0015&0.0016 &0.0009\\
              & MCSD        & 0.2252 & 0.1701 &0.1275&0.0954&0.0871 &0.0723\\
              & AASD       & 0.2148 & 0.1750 &0.1242&0.1063&0.0852&0.0744\\
              & MSE        &(0.0510)&(0.0291)&(0.0167)&(0.0091)&(0.0076)&(0.0052)\\
              &CP           & 0.95      &   0.97    & 0.93    &  0.97     &  0.96     &    0.95          \\
\\
$\alpha_{2}=0.3$
              & ABIAS       & 0.1795& 0.1335& 0.1038&0.0836&0.0723&0.0489\\
              & MCSD        & 0.2204 & 0.1916 & 0.1389&0.1116&0.0810&0.0788\\
              & AASD       & 0.2760 & 0.2993 & 0.1815&0.1494&0.0923&0.0803\\
              & MSE        &(0.0808)&(0.0545)&(0.0301)&(0.0194)&(0.0118)&(0.0086)\\
              &CP           & 0.97      &   0.96      &  0.94    &  0.96    & 0.96      &   0.96           \\
\\
$\beta_{1}=0.5$
              & ABIAS       &0.1120  &0.1102 &0.0936&0.0695&0.0440 &0.0141\\
              & MCSD        & 0.2586 &0.2399 & 0.1437&0.1204&0.0935&0.0414\\
              & AASD       & 0.2697 & 0.2382 & 0.1499&0.1267&0.0915&0.0428\\
         & MSE        &(0.0794)&(0.0697)&(0.0294)&(0.0193)&(0.0107)&(0.0019)\\
              &CP           & 0.96      &   0.94      &  0.94  &  0.96    &  0.95    &  0.94            \\
\\
$\gamma_{1}=-0.2$
              & ABIAS       &0.1057  &0.0702 &0.0264&0.0013 &0.0035&0.0027\\
              & MCSD        &0.4048  & 0.3744 & 0.2379&0.2321&0.1710&0.1257\\
              & AASD       & 0.3858  & 0.3829 &0.2384 &0.2353&0.1666&0.1227\\
         & MSE        &(0.1750)&(0.1451)&(0.0573)&(0.0539)&(0.0293)&(0.0158)\\
              &CP           & 0.92      &   0.97      &   0.92  &  0.96    &     0.94       &  0.94            \\
\\
$\gamma_{2}=0.5$
              & ABIAS       & 0.0661 &0.0183 &0.0131 &0.0273 &0.0108 &0.0103 \\
              & MCSD        & 0.2401 &0.1561 & 0.1519 &0.0753 &0.0891&0.0458\\
              & AASD       & 0.2508 &0.1503  & 0.1528&0.0746&0.0905&0.0462\\
         & MSE        &(0.0620)&(0.0247)&(0.0232)&(0.0064)&(0.0080)&(0.0022)\\
              &CP           & 0.93      &   0.97     &  0.96    &  0.94  &    0.94    &   0.95          \\
\hline
\end{tabular}
\caption{ABIAS, MCSD, AASD and MSE of $\bsb\alpha$, $\vect{\beta}$ and $\vect{\gamma}$ for samples with $\pi_c=40\%$.
}
\label{tab2}
}
\end{table}

Table \ref{tab1} presents the estimated results of $\vect{\alpha}=(\alpha_1, \alpha_2)^T$, $\beta_1$, and $\vect{\gamma}=(\gamma_1, \gamma_2)^T$, based on samples with $\pi_c=70\%$, and Table \ref{tab2} presents the estimated results based on samples with $\pi_c=40\%$. In these two tables (and also other tables), $\pi_R$ denotes the right censoring proportion in the entire sample, including those from the cured proportion.

These results suggest that, for a given sample size, values of ABIAS, MCSD, AASD, and MSE decrease as the non-cured proportion increases. The coverage probabilities for MPL tend to be close to the 95\% nominal value. Comparison of MCSD against AASD reveals that the sandwich formula given in (\ref{asyvar}) is accurate for the variance of the $\vect{\alpha}$, $\beta_1$, and $\vect{\gamma}$ estimates, and the accuracy improves when the non-cured proportion or sample size increases.

By comparing Table \ref{tab1} and Table \ref{tab2}, we observe that, for the same sample size and non-cured proportion, the quantities of ABIAS, MCSD, AASD, and MSE decrease with the censoring proportion $\pi_c$. Tables \ref{tab3} and \ref{tab4} report the results for estimated baseline hazards. We observe that, in general, these quantities reduce with increased sample sizes, increased non-cured proportions, or decreased censoring proportions.

\begin{figure}
    \centering
    \includegraphics[width=5.5in,keepaspectratio]{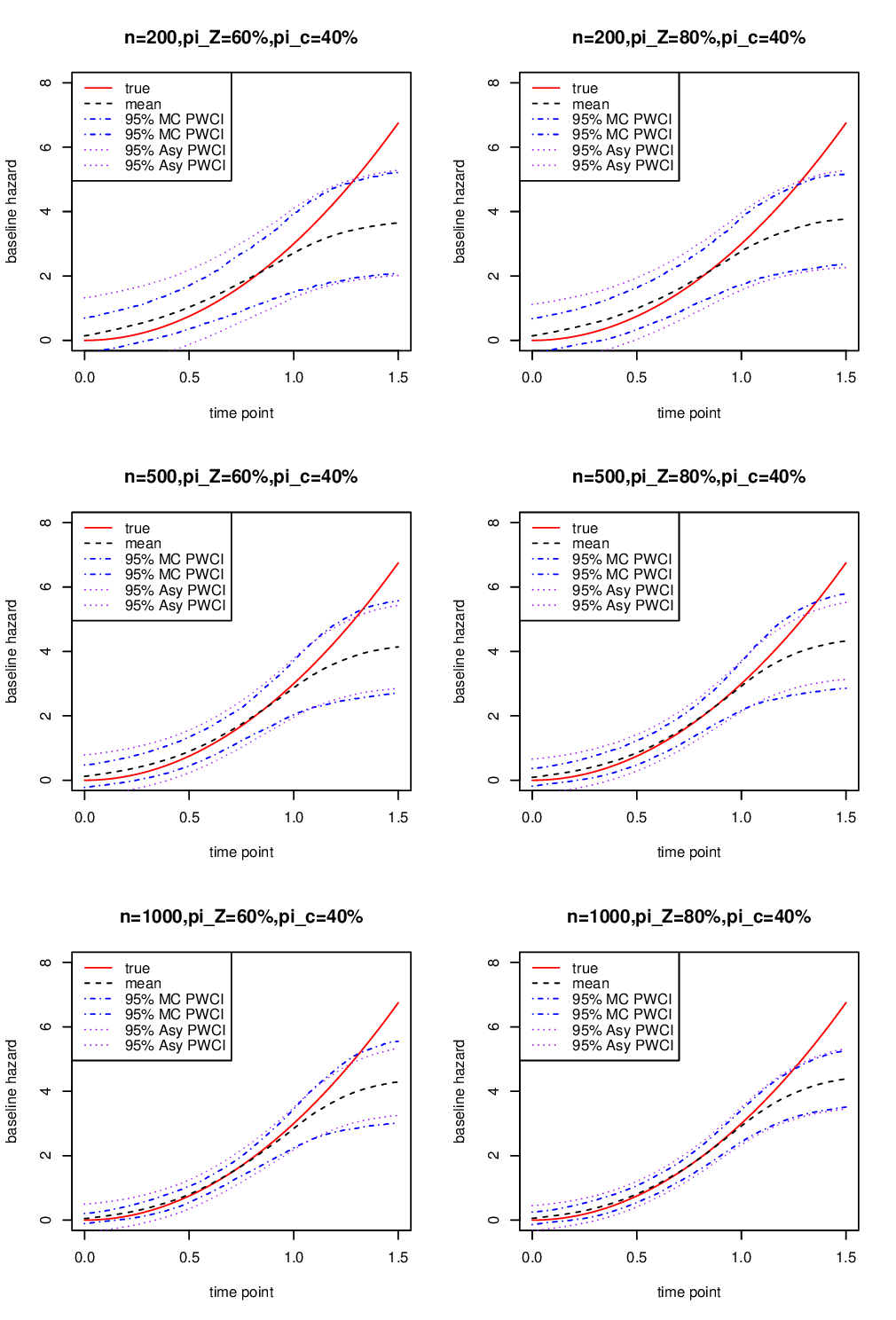}
    \caption{Plots of the true $h_{0}(t)$ (solid), the average MPL $h_{0}(t)$ estimates (dash), the 95\% Monte Carlo PWCI  (dot-dash), and the average 95\% asymptotic PWCI (dots)  .
}
    \label{fig1}
\end{figure}

\begin{figure}
    \centering
    \includegraphics[width=5.5in,keepaspectratio]{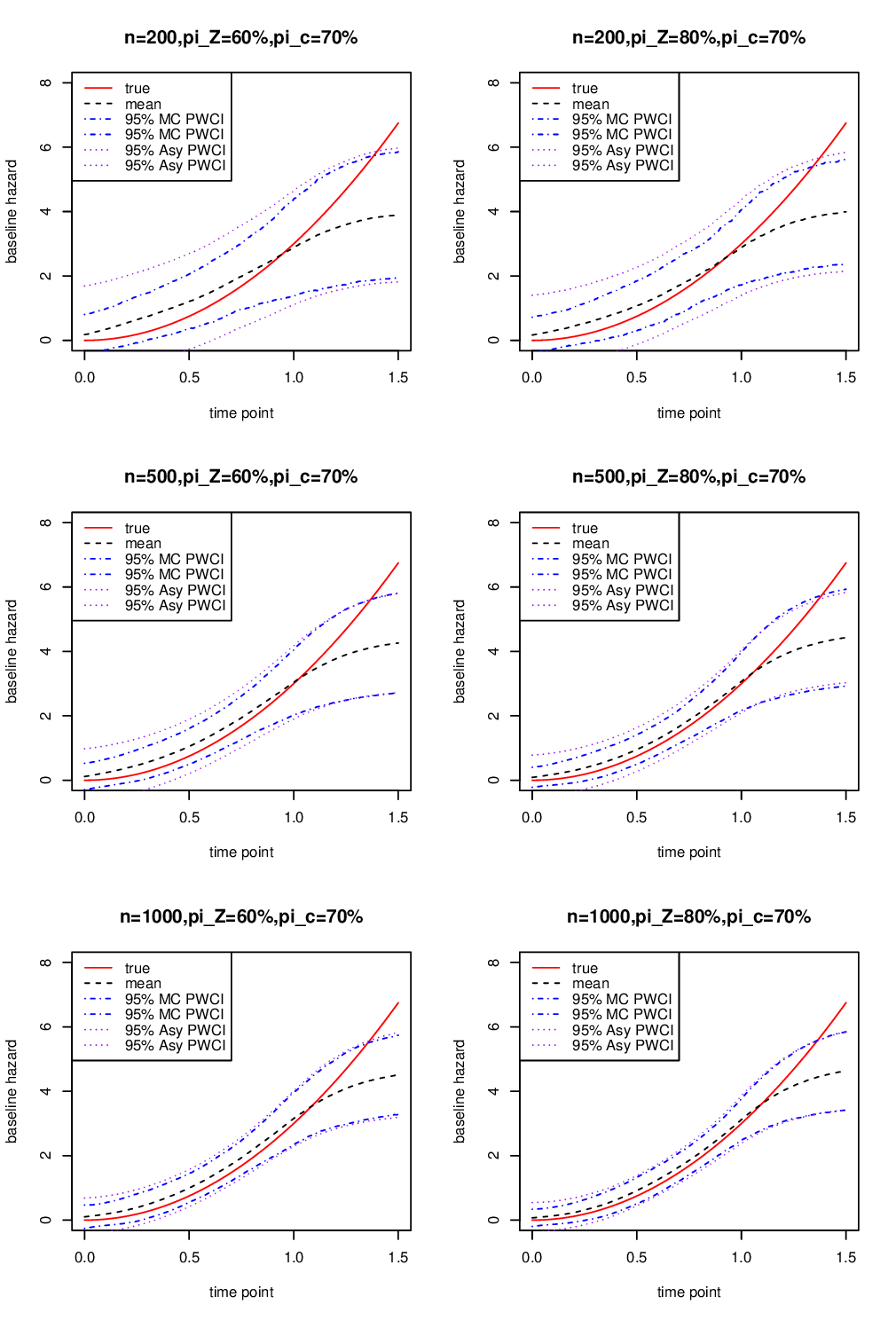}
    \caption{Plots of the true $h_{0}(t)$ (solid), the average MPL $h_{0}(t)$ estimates (dash), the 95\% Monte Carlo PWCI  (dot-dash), and the average 95\% asymptotic PWCI (dots)  .
}
    \label{fig1f}
\end{figure}

\begin{table}[th]
{\small
\centering
\begin{tabular}{l@{\hspace{-0.01in}}rr@{\hspace{0.001in}}rr@{\hspace{0.001in}}rr@{\hspace{0.001in}}rr@{\hspace{0.001in}}r@{\hspace{0.001in}}r}
\hline
\hline
& &\multicolumn{2}{c}{$n=200$}&\multicolumn{2}{c}{$n=500$}&\multicolumn{2}{c}{$n=1000$}\\
$\pi(\vect{z})$       &          &  60\%  &  80\%  & 60\%  & 80\% & 60\% &80\% \\
$n_{o}         $         &          &    2    &   2       &  3        &  3      &  4      & 4  \\

$\pi_R$       &          &   53\% &  37\%  &  53\% &  37\% &  53\%  &  37\%  \\
\hline
$h_0(t_1)$
              & ABIAS       & 0.4890& 0.3650&0.2960 &0.2073&0.2297&0.1402\\
              & MCSD        & 0.3989 & 0.3596 &0.2665&0.2253&0.2261&0.2006\\
              & AASD       & 0.7495 & 0.6045 &0.4240&0.3411&0.2889&0.2353\\
              & MSE        &(0.3983)&(0.2626)&(0.1587)&(0.0937)&(0.1039)&(0.0599)\\
              &CP           & 0.99      &   0.99      &   0.98  &  0.99    &  0.95    &   0.96           \\
\\
$h_0(t_2)$
              & ABIAS       &0.2993 & 0.1802 & 0.2013&0.1664&0.1471&0.1052\\
              & MCSD        & 0.5233 & 0.4486 & 0.3449&0.2887&0.2735&0.2261\\
              & AASD       & 0.8073 & 0.6518 & 0.4669&0.3812&0.3281&0.2713\\
              & MSE        &(0.3634)&(0.2337)&(0.1595)&(0.1110)&(0.0964)&(0.0622)\\
              &CP           & 0.99     &    0.98     &    0.98 &   0.97 &  0.96    &   0.96           \\
\\
$h_0(t_3)$
              & ABIAS       &0.4993 &0.4641&0.2609&0.1621&0.1046&0.0620\\
              & MCSD        & 0.8738 & 0.6972 & 0.6299&0.5898&0.4817&0.4309\\
              & AASD       & 0.9620 & 0.8250 & 0.6560&0.5793&0.5233&0.4629\\
              & MSE        &(1.0129)&(0.7015)&(0.4648)&(0.3741)&(0.2429)&(0.1896)\\
              &CP           & 0.97      &  0.96       &    0.97  & 0.97   &   0.95    &    0.96          \\
\\
$    h_0(t)    $  & AISE   & 1.8638  &  1.6119  &   1.1263     &   0.9333   & 0.8915   &    0.7418   \\
\hline
\end{tabular}
\caption{ABIAS, MCSD, AASD, MSE and AISE of 25th, 50th and 75th percentiles of  $\hat{h}_0(t)$ for samples with $\pi_c=70\%$.
}
\label{tab3}
}
\end{table}

\begin{table}[th]
{\small
\centering
\begin{tabular}{l@{\hspace{-0.01in}}rr@{\hspace{0.001in}}rr@{\hspace{0.001in}}rr@{\hspace{0.001in}}rr@{\hspace{0.001in}}r@{\hspace{0.001in}}r}
\hline
\hline
& &\multicolumn{2}{c}{$n=200$}&\multicolumn{2}{c}{$n=500$}&\multicolumn{2}{c}{$n=1000$}\\
$\pi(\vect{z})$       &          &  60\%  &  80\%  & 60\%  & 80\% & 60\% &80\% \\
$n_{o}         $         &          &    2    &   2       &  3        &  3      &  4      &  4  \\

$\pi_R$       &          &   47\% &  29\%  &  47\% &  29\% &  47\%  &  29\%  \\
\hline
$h_0(t_1)$
              & ABIAS       & 0.3101& 0.2691&0.1861&0.1435&0.0877&0.0863\\
              & MCSD        & 0.3143 & 0.3164 &0.2132&0.1807&0.1260&0.1130\\
              & AASD       & 0.5857 & 0.4833&0.3327&0.2341&0.1944&0.1889\\
              & MSE        &(0.1949)&(0.1725)&(0.0801)&(0.0532)&(0.0236)&(0.0202)\\
              &CP           & 0.97      &  0.97       &  0.97   &   0.96   &  0.95     &   0.96           \\
\\
$h_0(t_2)$
              & ABIAS       &0.1145 & 0.0936 &0.0541 &0.0353&0.0084&0.0046\\
              & MCSD        & 0.4556 & 0.3822 & 0.2612&0.2288&0.1575&0.1347\\
              & AASD       & 0.6253 & 0.5181 & 0.3678&0.3161&0.1951&0.1848\\
              & MSE        &(0.2207)&(0.1548)&(0.0711)&(0.0536)&(0.0249)&(0.0182)\\
              &CP           & 0.98      &  0.99       &  0.97  &  0.97     &  0.96    & 0.95             \\
\\
$h_0(t_3)$
              & ABIAS       & 0.5066 &0.5198 &0.2928&0.1550&0.1021&0.0511\\
              & MCSD        & 0.7061 & 0.6054 &0.5511 &0.5179&0.4233&0.3272\\
              & AASD       & 0.7551 & 0.6691 & 0.5362&0.4787&0.4041&0.3571\\
              & MSE        &(0.7552)&(0.6367)&(0.3894)&(0.2922)&(0.1896)&(0.1097)\\
              &CP           & 0.94      &  0.96       & 0.97     &   0.96   &  0.97    &      0.96        \\
\\
$  h_0(t)$   & AISE      & 1.8486     &   1.4872     &  1.1253     &   0.9128    &   0.7972  &     0.6469      \\
\hline
\end{tabular}
\caption{ABIAS, MCSD, AASD, MSE and AISE of 25th, 50th and 75th percentiles of  $\hat{h}_0(t)$ for samples with $\pi_c=40\%$.
}
\label{tab4}
}
\end{table}

Figure \ref{fig1} shows estimated baseline hazard plots, including the true $h_0(t)$ function, the mean of the MPL estimates of $h_0(t)$, and the $95\%$ piecewise confidence intervals (PWCIs) obtained from Monte Carlo standard errors as well as the average of the asymptotic standard errors. It can be observed that the mean MPL estimate of $h_0(t)$ becomes increasingly close to $h_0(t)$ as the sample size increases, the censoring proportion in the non-cured group decreases, or the non-cured proportion increases. These plots also indicate that both the Monte Carlo PWCIs and the average asymptotic PWCIs become narrower with a smaller censoring proportion in the susceptible group, a larger sample size, and/or a larger non-cured proportion, which suggests that the asymptotic variance of $\widehat{\vect{\theta}}$ is generally accurate.

\subsection{Application in a melanoma study}
Our proposed method was applied to analyze the time to first melanoma recurrence for 2209 patients diagnosed with melanoma in Australia. The dataset was provided by the Melanoma Institute Australia (MIA) and contained information on the date of follow-up visits, date of first recurrence diagnosis, recurrence and survival status at the last follow-up, baseline characteristics (age, sex, and body site of lesion), and pathological factors (Breslow thickness, ulceration, and mitoses count). Patients who experienced melanoma recurrence usually had their recurrence time interval-censored as the exact time of recurrence occurred were unknown to the doctors. We considered six categorical covariates in our model, namely, Breslow thickness ($\leq 0.6$ mm; $>0.6$ mm), tumor ulceration (yes; no), age group ($< 60$; $\geq 60$), sex (male; female), tumor mitoses (yes; no), and site of tumor (arm; head and neck; leg; trunk).

\begin{figure}
    \centering
    \includegraphics[width=4.2in,keepaspectratio]{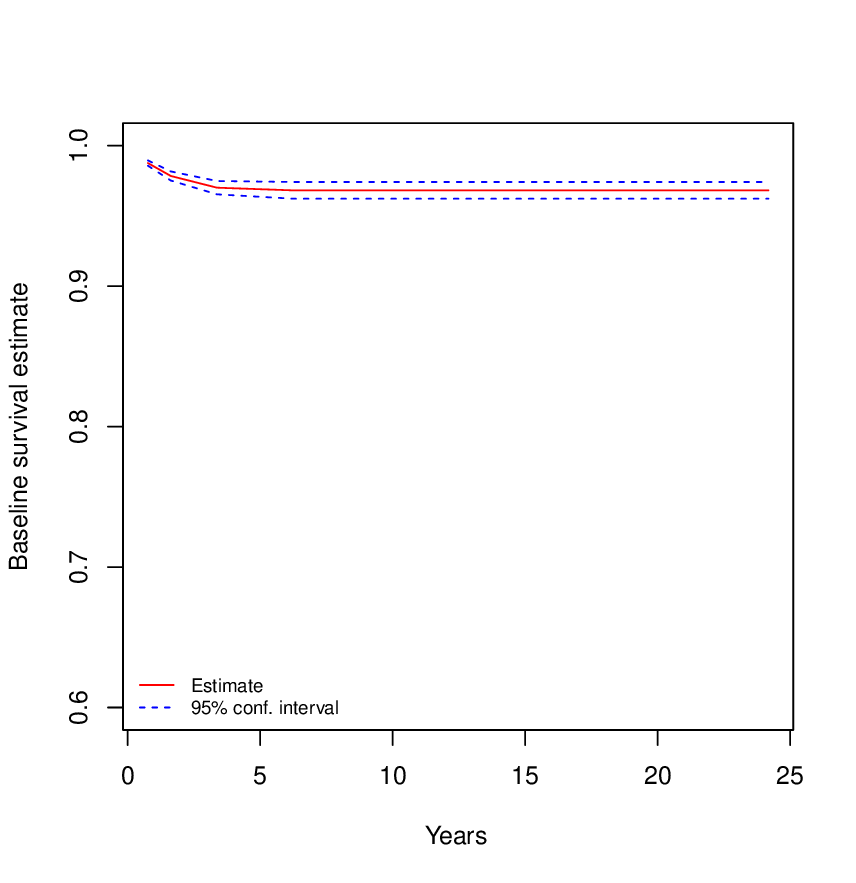} 
    \caption{Survival function estimate assuming no cured group. 
		}
    \label{fig2}
\end{figure}

Our analysis showed that only 6.74\% of the patients had experienced the first recurrence at some point during the follow-up study, while 93.26\% of the patients were right-censored, suggesting the presence of a cured fraction. We calculated the survival estimate based on this dataset assuming no cured group using the method proposed by \cite{LiMa19}. Figure \ref{fig2} shows the estimated survival function, when there is no cured fraction, becomes flat at around 0.97, indicating that some individuals may not experience melanoma recurrence. We then fitted a mixture cure AH model with all six covariates in the latency model and the incidence model.

\begin{table}[t]
\centering
\begin{tabular}{l*{6}{r}r}
\hline
              &\textit{Mixture-cure AH model} & & &\\
              &Covariates    & &          & \\
							\hline
              &\textbf{Incidence model}  &OR &  $p$-value & $95\%  \text{CI}$\\
              &Breslow thickness: $>0.6$mm & 0.960&  0.415  & (0.662, 1.392)\\
              &Ulceration: Yes       & 2.139&  0.015  & (1.077, 4.247)\\
              &Age: $>60$       & 0.419& $<0.000$  & (0.296, 0.594)\\
              &Sex: Male& 0.499  &$<0.000$ & (0.368, 0.677) \\
              &Mitoses: Yes &1.418  & 0.089&  (0.852, 2.361) \\
              &Body site: Head and Neck&0.216  &$<0.000$ &  (0.135 0.346) \\
              &Body site: Leg&0.087  & $<0.000$&  (0.056, 0.136) \\
              &Body site: Trunk&0.202  & $<0.000$&  (0.148, 0.277) \\
							\hline
              &\textbf{Latency model} &HD   & $p$-value & $95\%  \text{CI}$  \\
              &Breslow thickness: $>0.6$mm   & -0.289 &  0.110 & (-0.752, 0.174)\\
              &Ulceration: Yes       & -0.006 &  0.481 & (-0.258, 0.245)\\
              &Age: $>60$       & -0.033&  0.323 & (-0.173, 0.107)\\
              &Sex: Male& -0.212 & 0.020& (-0.415,  -0.010) \\
              &Mitoses: Yes &0.175  & 0.009&  (0.030, 0.319) \\
              &Body site: Head and Neck&0.138  & 0.102&  (-0.075, 0.352) \\
              &Body site: Leg&0.144  & 0.043&  (0.027 0.314) \\
              &Body site: Trunk&0.198  & 0.005&  (0.049 0.348) \\
  \hline
\end{tabular}
\caption{MPL mixture-cure AH model fitting results for the thin melanoma data.}
\label{tab5}
\end{table}

\begin{table}[t]
\centering
\begin{tabular}{l*{6}{r}r}
\hline
              &\textit{AH model, no cured fraction}\\
              &Covariates   & HD   & $p$-value & $95\%  \text{CI}$ \\
              &Breslow thickness: $>0.6$mm   & 0.0035 &  0.004 & (0.0009, 0.0060)\\
              &Ulceration: Yes       & 0.0087 &  0.043 & (0.0012, 0.0186)\\
              &Age: $>60$       & 0.0000&  0.500 & (-0.0025, 0.0025)\\
              &Sex: Male& 0.0004  & 0.394& (-0.0023, 0.0030) \\
              &Mitoses: Yes &0.0011 & 0.204&  (-0.0015, 0.0037) \\
              &Body site: Head and Neck&0.0018  & 0.131&  (-0.0013, 0.0049) \\
              &Body site: Leg&0.0000  & 0.500&  (-0.0031, 0.0031) \\
              &Body site: Trunk&0.0004  & 0.388&  (-0.0024, 0.0032) \\
\hline
\end{tabular}
\caption{MPL AH model fitting results for the thin melanoma data.}
\label{tab6}
\end{table}

\begin{figure}
    \centering
    \includegraphics[width=4.2in,keepaspectratio]{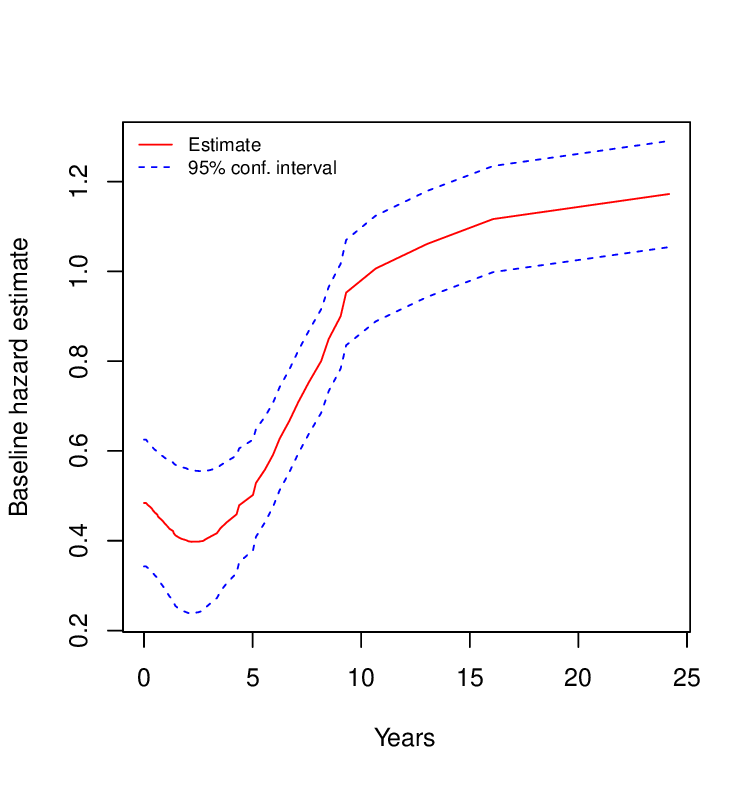}
    \caption{Baseline hazard estimate for the non-cured population, where a mixture cure AH is assumed.}
    \label{fig3}
\end{figure}

\begin{figure}
    \centering
    \includegraphics[width=4.2in,keepaspectratio]{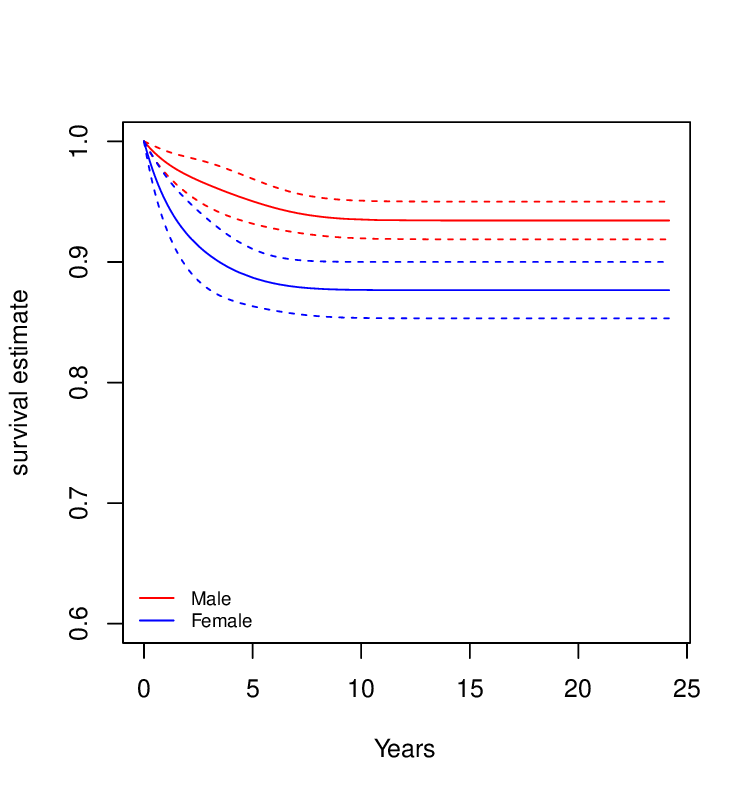}
    \caption{Mixture cure predictive survival functions for Male vs Female at the 
population level.}
    \label{fig4}
\end{figure}

\begin{figure}
    \centering
    \includegraphics[width=4.2in,keepaspectratio]{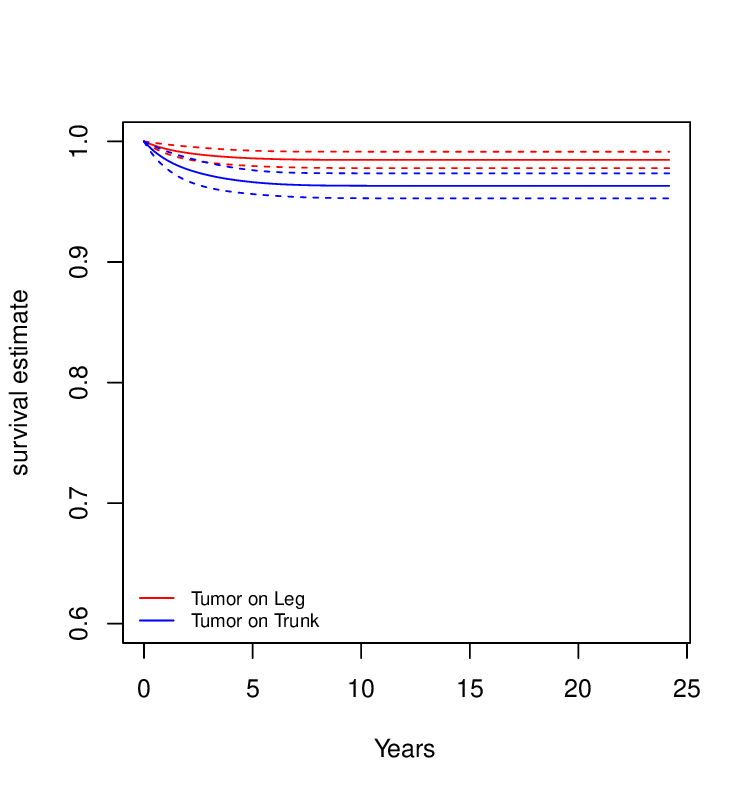}
    \caption{Mixture cure predictive survival function for Tumor on Leg vs Tumor on Trunk at the 
population level.}
    \label{fig5}
\end{figure}

The top half of Table \ref{tab5} shows the estimates for the odds ratio (OR) in the incidence logistic regression model, and the bottom half of this table displays the hazard differences (HD) in the latency additive hazards model, along with $p$-values and 95\% confidence intervals (95\% CI). The results from the incidence model demonstrate that the odds of thin melanoma recurrence are lower for patients who had a tumor on the head and neck, leg, or trunk than for those who had a tumor on the arm (the reference category). Male patients or patients aged over 60 have a lower odds ratio of being classified into the non-cured sub-population. The odds increase in patients with tumor ulceration. Given in the melanoma recurrence sub-population, males have a lower risk of recurrence than females and patients with mitoses have a higher risk. 
A tumor on the leg or trunk instead of the arm also increases the risk of melanoma recurrence.

Table \ref{tab6} contains the fitting results without assuming a cured fraction, using the method given by \cite{LiMa19}. The results appear very different from Table \ref{tab5}. Particularly, assuming there is no cured fraction, patients with ulceration or Breslow thickness larger than 0.6 mm have a higher risk of melanoma recurrence.

Figure \ref{fig3} exhibits the baseline hazard estimate with 95\% point-wise confidence intervals for the melanoma recurrence sub-population in the mixture-cure AH model. There is a notably decreasing trend in the risk of tumor recurrence from year 0 to year 4, after which the risk of recurrence increases sharply until year 10. From year 10 and year 25, a small increasing presents although the point-wise confidence intervals are becoming wider during this period.

Our proposed method is capable of making predictions and computes easily the mixture cure survival values using logistic and additive hazards regression parameter estimates and the conditional baseline hazard estimates. For instance, the estimated probability of a female patient having no recurrence for 3 years is 0.906 with 95\% confidence interval $(0.877, 0.934)$, with all other covariates being fixed at mean values,  while this probability is increased to 0.964 with 95\% confidence interval $(0.945, 0.982)$ for a male patient.  Consider a patient who has a tumor on leg, the estimated probability of no recurrence for 6 years is 0.985 with 95\% confidence interval $(0.979, 0.992)$.  And this estimate is 0.965 with 95\% confidence interval $(0.955, 0.975)$ for a patient who has a tumor on trunk.

We can also estimate the entire predictive survival probability and 95\% point-wise confidence interval (CI) based on the mixture cure model. Figure \ref{fig4} displays mixture cure predictive survival functions demonstrating the difference of recurrence risks between male patients and female patients at the population level.  It clearly illustrates that, with all other covariates equal to their sample mean values,  male patients have higher probability 
of no recurrence at any time $t$ than female patients.  Figure \ref{fig5} presents the mixture cure predictive survival functions at the population level to compare the recurrence risk of tumor on leg with the risk of tumor on trunk.  It is clear that patients with tumor on trunk are more likely to have recurrence 
than those with tumor on leg, with all other covariates equal to their sample mean values.

\section{Conclusion} \label{sec:conc}
This paper develops a maximum penalized log-likelihood method to “fit the mixture cure 
additive hazard model with partly interval-censored failure time data, where a penalty function is included to ensure smoothness of the estimated baseline hazard function, as well as to achieve numerical stability of the algorithm.

In the estimation procedure, the baseline hazard is approximated using basis functions, and the estimates of the hazard for each individual and the baseline hazard are constrained to be non-negative. The constrained optimization computation is conducted by the primal-dual interior-point algorithm, which provides simultaneously the MPL estimates of the baseline hazard and the regression coefficients.

$\sqrt{n}$-consistency and asymptotic normality of the MPL estimates are proved, and the asymptotic standard deviations are generally accurate as assessed from the simulation study. Although we assume independent censoring in developing the MPL method, we can extend our method to allow dependent censoring by copulas similar to \cite{XuMaCoBr18}. This will be our future research topic.


\end{document}